# Strong light-matter interaction with self-hybridized bound states in the continuum in monolithic van der Waals metasurfaces

*Thomas Weber[1], Lucca Kühner[1], Luca Sortino[1], Amine Ben Mhenni[2], Nathan P. Wilson[2], Julius Kühne[1],*

*Jonathan J. Finley[2], Stefan A. Maier[3,4,1] and Andreas Tittl[1,\*]*

1. Ludwig-Maximilians-Universität München, Faculty of Physics, Chair in Hybrid Nanosystems, 80539 Munich, Germany
2. Walter Schottky Institut, Department of Physics and MCQST, Technische Universität München, 85748 Garching, Germany
3. School of Physics and Astronomy, Monash University, Clayton, Victoria 3800, Australia.
4. Department of Physics, Imperial College London, London SW7 2AZ, UK

*e-mail: andreas.tittl@physik.uni-muenchen.de

**Abstract**

Photonic bound states in the continuum (BICs) are a standout nanophotonic platform for strong light-matter coupling with transition metal dichalcogenides (TMDCs), but have so far mostly been employed as all-dielectric metasurfaces with adjacent TMDC layers, incurring limitations related to strain, mode overlap, and material integration. In this work, we experimentally demonstrate for the first time asymmetry-dependent BIC resonances in 2D arrays of monolithic metasurfaces composed solely of the nanostructured bulk TMDC $WS_2$ with BIC modes exhibiting sharp and tailored linewidths, ideal for selectively enhancing light-matter interactions. Geometrical variation enables the tuning of the BIC resonances across the exciton resonance in bulk $WS_2$, revealing the strong-coupling regime with an anti-crossing pattern and a Rabi splitting of 116 meV. The precise control over the radiative loss channel provided by the BIC concept is harnessed to tailor the Rabi splitting via a geometrical asymmetry parameter of the metasurface. Crucially, the coupling strength itself can be controlled and is shown to be independent of material-intrinsic losses. Our BIC-driven monolithic metasurface platform can readily incorporate other TMDCs or excitonic materials to deliver previously unavailable fundamental insights and practical device concepts for polaritonic applications.

**Introduction**

Understanding and maximizing the interaction between light and matter in nanoscale materials is one of the core goals of nanophotonics. A broad range of resonant nanosystems have been investigated for the confinement and control of the electromagnetic energy in sub-wavelength volumes, leading to breakthroughs in light harvesting[1], optical waveguiding[2], and emission control[3]. Coupling light to electronic excitations in solid state materials is of particular interest because of the creation of hybridized photonic and electronic states, called polaritons, showing exciting properties such as Bose-Einstein condensation[4] and superfluidity[5] with potential for applications in low-threshold semiconductor lasers[6], photocatalytic enhancement[7], and quantum computing[8]. The driving force for research on excitonic coupling has been reaching the strong light-matter coupling regime, where irreversible and coherent exchange of energy between photons and excitons is observed. Transition metal dichalcogenides (TMDCs) are a class of van der Waals materials and promising candidates for reaching the strong light-matter coupling regime. They host strongly bounded excitons stable up to room temperature, owing to large binding energies > 200 meV[9], coupled valley and spin degrees of freedom, and broad tunability through strain, dielectric environment and the DC Stark effect[10–12]. Because of their strong excitonic oscillator strengths, TMDCs show pronounced excitonic resonances even in tens to hundreds nm thick films[13], rendering this class of materials compatible with established architectures of flat-optics devices to study strong light-matter interaction.

In general, reaching the strong coupling regime in nanophotonic systems requires high electromagnetic field intensities, confinement of light into small modal volumes, and a maximized overlap between the optical modes of the cavity and the excitonic resonances in the material. To date, strong coupling has been realized with TMDC monolayers in a variety of photonic systems, including plasmonic nanoantennas[14,15], photonic crystals[16,17], and distributed Bragg reflector microcavities[18], as well as in self-hybridizing structures such as bulk TMDC slabs acting as Fabry-Perot cavity[19] and single nanostructures supporting Mie resonances[20].

Recently, photonic bound states in the continuum (BICs) have been shown to provide excellent control over engineering light-matter interactions, owing to their strongly enhanced near fields, high quality factors, and broad resonance tunability via variation of their geometrical parameters. Thanks to these versatile properties, BICs have established themselves as essential new building blocks for nanophotonic vibrational spectroscopy[21,22], light guiding[23], harmonic generation[24], and lasing[25]. Symmetry-protected metasurfaces[26] have emerged as a prominent platform to realize BIC resonances for light-matter coupling applications, owing to their high optical signal contrasts, experimental robustness[27], and ease of optical measurements requiring only common brightfield microscopy setups. Significantly, the BIC metasurface concept provides direct control over the radiative decay rate (and therefore the linewidth) of the resonances via the asymmetry of the constituent unit cells[28], enabling additional degrees of freedom for tailoring light-matter interactions. Currently, many implementations and proposals of excitonic strong coupling with BICs rely on placing TMDC layers adjacent to BIC metasurfaces composed of traditional all-dielectric materials such as Si, $TiO_2$, $Si_3N_4$, or $Ta_2O_5$[17,29–32]. However, this approach requires error-prone fabrication techniques leading to increased system complexity due to the direct contact between different material systems and inhomogeneities introduced during the TMDC transfer process. In such systems, light-matter interaction occurs only via the evanescent fields around the BIC resonators, which decreases the coupling strength, especially when incorporating commonly used buffer layers for TMDCs, such as hexagonal boron nitride (hBN).

Obtaining a monolithic TMDC BIC platform, which combines photonic cavity and excitonic material in the same nanostructured system is therefore highly desirable, however, an experimental demonstration is still lacking. Here, we realize strong coupling in monolithic $WS_2$ metasurfaces based on symmetry-protected BICs. The metasurface platform consists of arrays of rod-type BIC unit cells, where the asymmetry of the structure is controlled via the length difference $\Delta L$ of the rods, offering direct control over the resonance linewidth. We experimentally demonstrate pronounced and tunable BIC resonances in $WS_2$ metasurfaces fabricated via mechanical exfoliation, electron beam lithography, and reactive ion etching (see Methods for more details), with quality factors approaching $Q = 370$, at spectral positions away from the exciton. By varying the lateral size of the metasurface unit cells, the BIC resonances are tuned across the WS$_2$ exciton, which reveals an anti-crossing pattern with a Rabi splitting of 116 meV at ambient conditions, which is three times larger than the linewidths of the underlying excitonic and photonic resonances, propelling the system well into the strong-coupling regime. The optical performance of the WS$_2$ metasurface is maintained at cryogenic temperatures, showing a characteristic reduction of the Rabi splitting associated with a decrease of the excitonic oscillator strength. Using the versatile resonance control afforded by the BIC concept, our experiments reveal a clear increase of the Rabi splitting with lower asymmetries, which we attribute to a complex interplay between radiative quality factor and mode volume. Our results demonstrate that monolithic WS$_2$ metasurfaces can provide a versatile toolkit for studying light-matter interactions in the strong coupling regime with excellent control over the resonances. Monolithic BIC metasurface platforms can be designed and fabricated for a multitude of other TMDCs or excitonic materials, setting the stage for novel applications of exciton-polaritons in dielectric nanophotonic architectures.

**Results**

**Monolithic metasurface design**

As representative of the class of TMDCs, WS$_2$ was chosen for the monolithic BIC metasurface realization because of its spectrally isolated exciton at 629 nm for the bulk material with the largest oscillator strength compared to other TMDCs such as MoS$_2$, MoSe$_2$ or WSe$_2$[13] and high refractive index (n = 4.1 at $\lambda$ = 800 nm). Our design adopts a rod-type BIC unit cell geometry (**Figure 1a**), which provides precise control over the asymmetry via the length difference $\Delta L_0$, high signal modulation, and a spectrally clean mode structure. The WS$_2$ metasurfaces were numerically optimized to simultaneously obtain spectral overlap of the BIC resonances with the exciton, strong near field enhancements, and compatibility with the nanofabrication process. In this work, the basic geometrical unit cell parameters are fixed as: periodicity $P_0$ = 340 nm, base rod length $L_0$ = 266 nm, and rod width $W_0$ = 90 nm. The center points of the rods are placed at the unit cell coordinates (x, y) = (0, $P$/4) and (0, 3$P$/4), respectively. Tunability of the resonance position over the spectral location of the exciton is realized by introducing a multiplicative scaling factor $S$, which scales the in-plane geometrical parameters according to $P = S \cdot P_0$, $L = S \cdot L_0$, $\Delta L = S \cdot \Delta L_0$ and $W = S \cdot W_0$. In order to capture the optical response of the $WS_2$ BIC metasurface with and without the influence of the exciton (i.e., in the coupled and uncoupled regimes), we model the refractive index of $WS_2$ as a parametric Tauc-Lorentz dielectric (**Figure 1b**, see Methods). We demonstrate the existence of a BIC on the low energy side of the exciton with full-wave simulations and show the capability to spectrally tune the BIC via the geometric scaling factor $S$ (**Figure 1c**) and to control the resonance linewidth with a varying asymmetry parameter $\Delta L_0$ (**Figure 1d**). At resonance, the electric near field $\varepsilon E$, where $\varepsilon$ is the permittivity of $WS_2$ or air, is concentrated inside the structure (**Figure 1e**), leading to an enhancement of the light-matter interaction. To gain deeper insights into the loss channels associated with the BIC modes, we employ a temporal coupled mode theory (TCMT) model to fit the simulated transmittance spectra. This approach allows us to decompose the quality factors of the BICs into their radiative and intrinsic parts, where the radiative contribution shows the expected characteristic inverse square law with asymmetry. Due to small losses in the material the total Q factor is limited by the intrinsic quality factor, which leads to a deviation of the inverse square dependence. Additionally, the BIC-driven light-matter interaction is governed by critical coupling[33], where the highest field enhancement is achieved when the radiative and intrinsic Q factors match (**Figure 1e,f**).

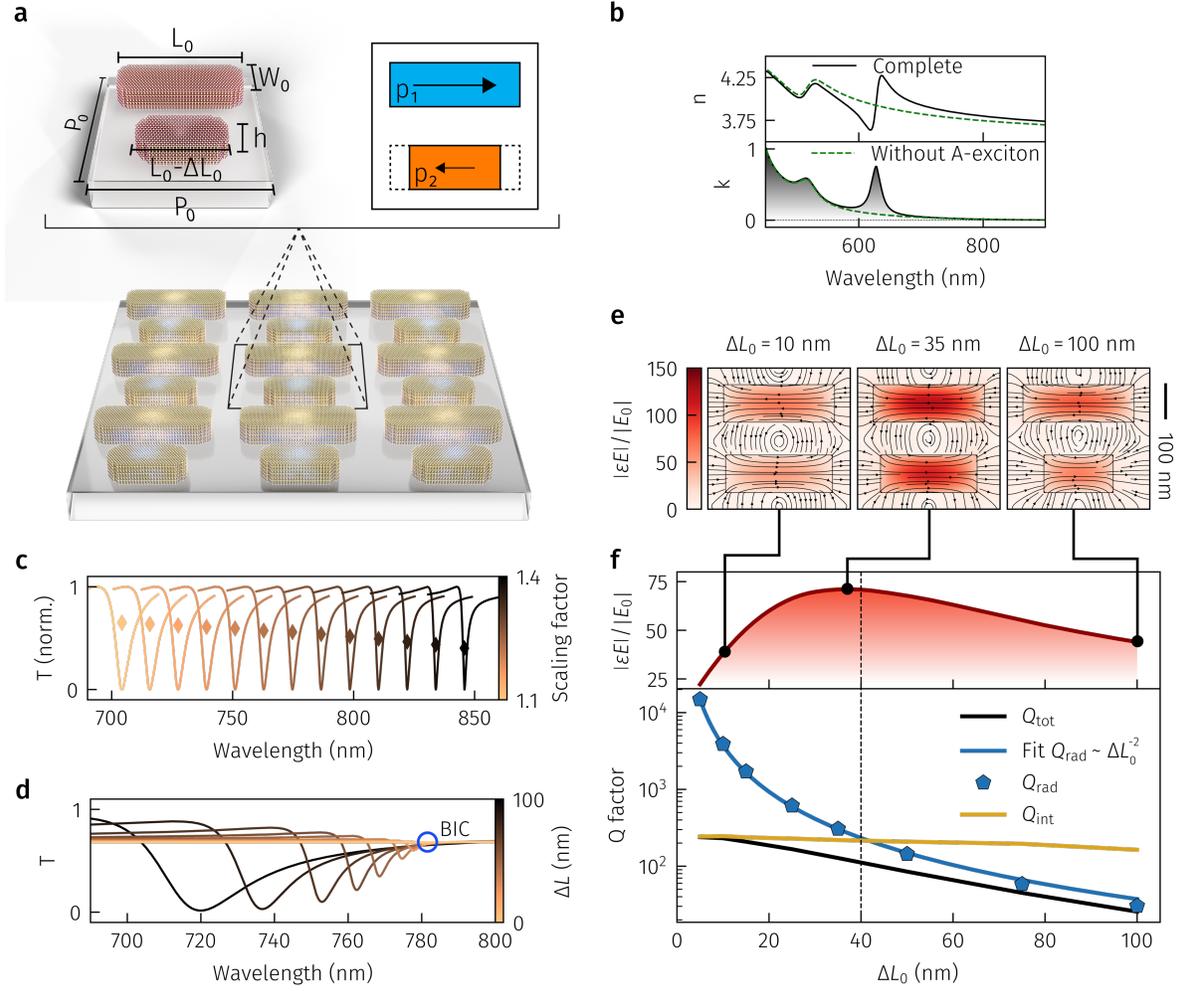

**Figure 1: Bound states in the continuum in bulk WS$_2$ metasurfaces. a**, Monolithic WS$_2$ metasurface supporting symmetry protected bound states in the continuum based on a rod-type BIC unit cell showing the BIC working principle of opposing dipoles. **b**, Parametric Tauc-Lorentz material of WS$_2$ showing the in-plane complex refractive index with and without the exciton. **c**, Simulated transmittance spectra of BIC resonances on the low energy side of the exciton tuned via scaling of the in-plane geometric parameters. The markers indicate the modulation of the transmittance signal, which increases for higher wavelengths because of a decrease of intrinsic losses. **d**, Simulated transmittance spectra of BICs for different asymmetries $\Delta L_0$. Smaller asymmetries lead to a spectral redshift and a reduction of the linewidth. For symmetric structures the resonance vanishes as the quasi-BIC turns into a true BIC, which is not excitable from the far field. **e-f**, Electric field enhancements and quality factors for different asymmetry parameters $\Delta L_0$. Due to intrinsic losses of the material, the maximal field enhancement is achieved when both intrinsic and radiative damping rates of the BIC mode are matched. The radiative quality factor follows the expected inverse square dependence of a BIC. The intrinsic Q factor shows a slight increase for lower asymmetries due to a slightly larger extinction value of the blue-shifted BIC. The total Q factor follows the radiative Q factor for large asymmetries and is dominated by the intrinsic Q factor for small asymmetries, which sets an upper limit.
6

**Metasurface fabrication and optical characterization**

As seen in **Figure 2a**, the fabrication process begins with the mechanical exfoliation of bulk $WS_2$ flakes onto fused silica substrates. By controlling exfoliation parameters, such as the heat treatment temperature, exfoliation time and applied pressure, we optimized the yield of the flakes with thicknesses ranging from 30 to 50 nm to achieve an adequate area to host a multitude of BIC metasurfaces. The exfoliated $WS_2$ flakes were then patterned with the BIC design via electron-beam lithography (EBL) and reactive-ion etching (RIE) (see Methods). The resulting metasurface patches have a footprint of 35x35 $\mu m^2$, which ensures a sufficient number of more than 50x50 unit cells for each scaling factor – a necessity to excite the collective BIC resonance with stable Q factor[34]. The transmittance spectra of the fabricated structures were measured in an optical spectroscopy setup, taking advantage of the large footprint of the metasurfaces to collect light only in the center of the arrays, which mitigates edge effects and allows for optimal signal modulations. SEM micrographs for different asymmetry parameters reveal the accurate reproduction of the metasurface design with high uniformity and surface quality (**Figure 2b**). We elucidate the emergence of the BIC mode from the symmetric case by gradually increasing the asymmetry parameter $\Delta L_0$. The corresponding transmittance spectra (**Figure 2c**) clearly show a spectral blueshift and linewidth broadening with increasing asymmetry, and the extracted Q factors range from 20 to 370, with highest values found for the lowest asymmetries (**Figure 2d**), confirming the nature of the resonances as symmetry-protected BICs. Overall, we see an excellent agreement between experimental and simulated results, although the radiative Q factor slightly deviates from the characteristic inverse square dependence of the BIC due to fabrication tolerances for low asymmetries, which we attribute to additional parasitic loss channels associated with fabrication imperfections.

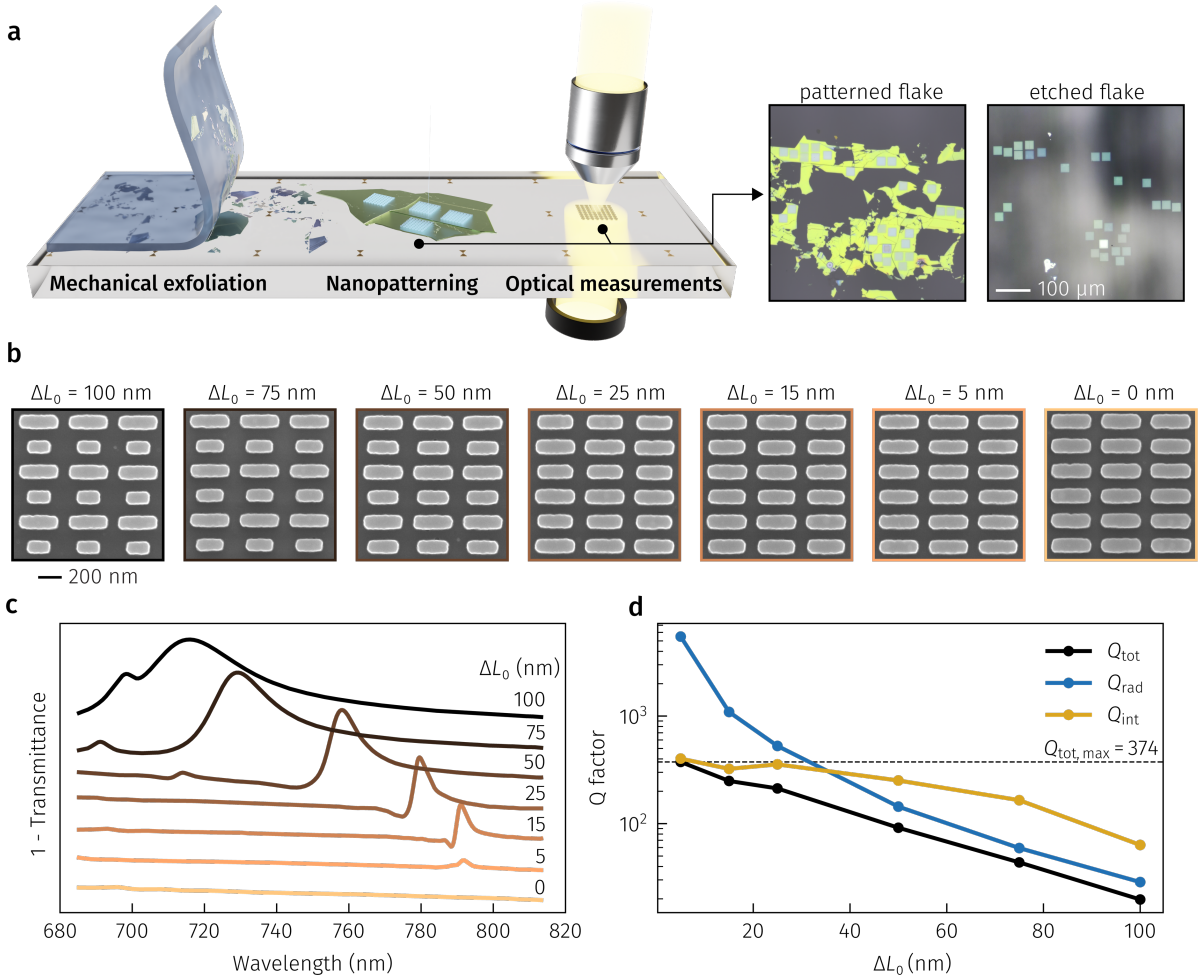

**Figure 2: Experimental realization of bulk WS$_2$ BIC metasurfaces. a**, Sketch of the experimental process including exfoliation of WS$_2$, nanopatterning via EBL and RIE, and optical far-field spectroscopy. **b**, SEM micrographs of fabricated WS$_2$ metasurfaces with a scaling factor of $S = 1.25$ with different asymmetry parameters $\Delta L_0$ and a uniform thickness of 80 nm. **c**, Transmittance spectra of WS$_2$ metasurfaces (shifted for visiblility) for different $\Delta L_0$, showing the formation of the quasi-BIC modes. To show the uncoupled BIC modes, resonances were placed spectrally separated from the WS$_2$ exciton. **d**, Resonance Q factors extracted from the spectra in panel (c) showing a maximum Q factor of 374.

**BIC-driven intrinsic strong coupling**

We now leverage our monolithic TMDC BIC metasurface platform for strong light-matter coupling by tuning the associated resonances over the WS$_2$ exciton wavelength at 629 nm (1.971 eV). The spectral tuning is achieved by fabricating a total of 11 metasurface patches with an asymmetry parameter of $\Delta L_0 = 80$ nm with scaling factors between S = 0.9 and S = 1.2 from a single WS$_2$ flake with a height of $h = 45$ nm. Representative transmittance spectra of the different metasurface patches show a clear anti-crossing behavior, which is a necessary feature of the strong coupling regime (**Figure 3a**). The full energy dispersion of the system is shown in Figure 3b and is plotted against the energy to facilitate the following analysis. The energies of the upper and lower polariton branches can be calculated from

$$\omega_{\pm} = \frac{\omega_{BIC} + \omega_{Ex}}{2} + \frac{i(\gamma_{BIC} + \gamma_{Ex})}{2} \pm \sqrt{g^2 - \frac{1}{4}(\gamma_{BIC} - \gamma_{Ex} + i(\omega_{BIC} - \omega_{Ex}))^2}, \quad (1)$$

where $\omega_{Ex}$ and $\gamma_{Ex}$ indicate the frequency and the exciton damping rate, $\omega_{BIC}$ and $\gamma_{BIC}$ are the resonance frequency and damping rate of the BIC, respectively, and $g$ is the coupling strength between the two coupled systems (see Supplementary Note 2). The corresponding Rabi splitting is defined as $\Omega_R = 2\sqrt{g^2 - (\gamma_{BIC} - \gamma_{Ex})^2/4}$. The properties of the exciton are extracted from transmission measurements on a symmetric metasurface patch without the presence of a BIC mode, yielding $\hbar\omega_{Ex} = 1.971$ eV and $\hbar\gamma_{Ex} = 36$ meV. The linewidth of the BIC at the wavelength of the exciton is obtained from simulations as $\hbar\gamma_{BIC} = 30$ meV (**Figure 3c**). Based on the values of these parameters, we can now fit Equation (1) to the experimentally obtained energies of the polariton branches to obtain a Rabi splitting of $(116 \pm 4)$ meV. We evaluate two criteria of strong coupling[31,35]

$$c_1 = \Omega_R/(\gamma_{BIC} + \gamma_{Ex}) > 1 \quad (2)$$

$$c_2 = g/\sqrt{(\gamma_{BIC}^2 + \gamma_{Ex}^2)/2} > 1 \quad (3)$$

which yield $c_1 = 1.75$ and $c_2 = 1.74$, respectively, verifying that our monolithic metasurface concept indeed reaches a stable point in the strong-coupling regime (see **Figure S3**).

Furthermore, we investigate the influence of temperature on the coupling process by measuring the transmittance spectra of the WS$_2$ BIC metasurface at different temperatures in a cryostat. The optical performance of the metasurface is maintained even at cryogenic temperatures, and comparing the experiments at room temperature and 5 K, the expected spectral blueshift of the exciton[36] is visible (**Figure S5a**). Although the Rabi splitting shows a small declining trend with decreasing temperatures (**Figure S5b**), which we ascribe to the reduction of the oscillator strength for the phonon-mediated momentum indirect transition of the bright exciton, strong coupling is maintained down to 5 K.

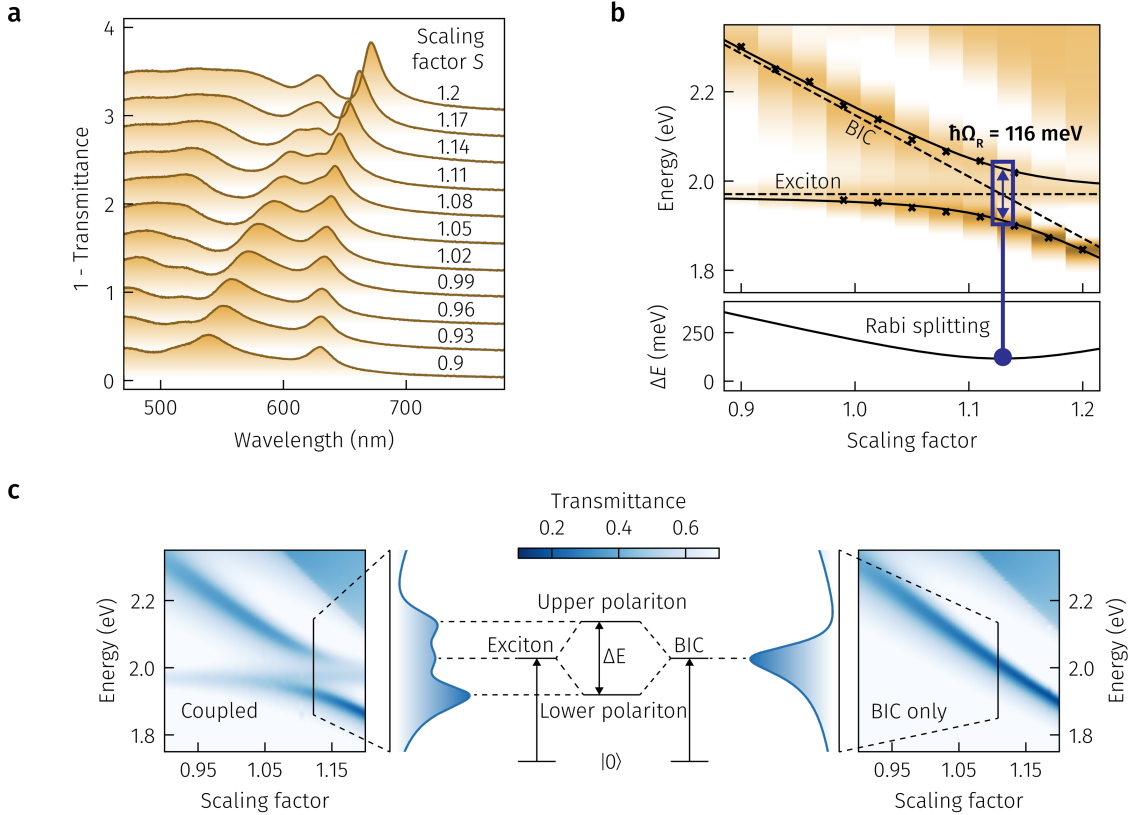

**Figure 3: Strong coupling in WS$_2$ BIC metasurfaces. a**, Experimental transmittance spectra of WS$_2$ BIC metasurfaces with an asymmetry parameter $\Delta L_0 = 80$ nm for different scaling factors show a characteristic anti-crossing mode pattern close to the WS$_2$ exciton. **b**, The energy dispersion fit of both polariton branches reveals a Rabi splitting of 116 meV at room temperature. **c,** Simulations where the WS$_2$ BIC mode is continuously tuned over the spectral location of the WS$_2$ exciton by varying the scaling factor using dielectric functions from the Tauc-Lorentz Model with and without the exciton. By aligning the spectral location of BIC and exciton both modes hybridize into polaritonic branches with higher and lower energies compared to their uncoupled ground states.

**Tailored strong coupling**

To study the influence of the BIC quality factor on the Rabi splitting, we simulate and fabricate metasurfaces with different asymmetry factors $\Delta L_0$ with a height of 33 nm. For higher asymmetries, the BIC resonances exhibit higher modulation contrasts combined with broader linewidths (**Figure 4a,b**). On the other hand, we observe increased values of the Rabi splitting for lower asymmetries (**Figure 4c**). This unique tuning behavior provides a new mechanism for tailoring the Rabi splitting using simple structural modification, and can thus enable the design of metasurface geometries with optimal ratio of resonance modulation to Rabi splitting in the strong-coupling regime (**Figure S4**). For the lowest asymmetry of $\Delta L_0 = 20$ nm, the BIC resonances are strongly damped by the intrinsic losses of $WS_2$, and we can no longer resolve a clear peak splitting (grey shaded area in Figure 4c).

To gain further insights on this scaling behavior, we plot the Rabi splitting against the radiative Q factors of the BIC resonances without exciton from simulations and observe that both the Rabi splitting and the coupling constant $g$ undergo a steep increase at low Q factors and saturate at higher Q factors (**Figure 4d**). The observed trend is consistent with predictions from the literature, where the saturation of the Rabi splitting is attributed to the saturation of the electric field enhancement[37,38]. However, in comparison to idealized lossless dielectric BIC-driven metasurfaces, where the field enhancement scales with the square root of the Q factor (**Figure S7i**) and thus shows a saturating behavior, the field enhancement of realistically modeled monolithic $WS_2$ BIC metasurfaces is governed by critical coupling, which is determined by background losses in the material. At the spectral position of the exciton, the maximal field enhancement is achieved for an asymmetry parameter $\Delta L_0 = 110$ nm (**Figure 4c**) and is showing an upward trend, which contrasts strongly with the downward trend of the Rabi splitting, implying that the Rabi splitting is independent of the electric near field enhancement, which is also supported by additional simulations with lossless and lossy materials, showing identical Rabi splitting (**Figure S7c,f**).

This independence of field enhancement is consistent with the common definition for the coupling strength $g$[39]

$$g \sim \boldsymbol{\mu} \cdot \mathbf{E} \sim \sqrt{\frac{N}{V_{\text{eff}}}}, \quad (4)$$

where $\boldsymbol{\mu}$ is the collective dipole moment of the exciton, $\mathbf{E}$ the local electric field strength, $N$ the number of excitons participating in the coupling process and $V_{\text{eff}}$ the effective mode volume. We calculated the effective mode volume, which follows the decreasing trend of the geometric volume, when increasing the asymmetry (Supplementary Note 4, **Figure S8**), which would imply an increase of coupling strength. However, due to the reduced volume, fewer excitons partake in the coupling process, which effectively counteracts the effects of the reduction of mode volume, leading to a constant exciton density. Thus, the coupling strength $g$ should be independent of geometric variations, which cause changes in the electric field strength. As mentioned by Tserkezis et al.[40], the above definition of the coupling strength is only valid for low-loss and nonradiative cavities. Our BIC based platform facilitates the flexible control of the degree of radiative losses, which makes it necessary to study the influence of largely varying Q factors on the strong coupling process in more detail.

To further understand the intrinsic coupling behavior of our monolithic $WS_2$ BIC metasurface, we separately study the radiative and intrinsic loss channels that affect the total BIC Q factor. For this purpose, we use a simplified Lorentzian material representing a narrower exciton with a linewidth that is between the narrowest and the broadest BIC linewidths achievable with the given geometric parameters. When we change the Q factor by adding intrinsic losses to the system we observe a constant coupling strength $g$ and a Rabi splitting peaking at the maximum value $2g$ for matching linewidths of BIC and exciton (**Figure S6a,b**) and the decrease of Rabi splitting for low Q factors is solely caused by weakening the strong coupling conditions moving towards the weak coupling regime (**Figure S6c**).

A different picture emerges when we change the BIC quality factor by changing radiative losses through a variation of the asymmetry parameter $\Delta L_0$. Both the coupling strength and Rabi splitting follow the saturating trend described above, independent of intrinsic losses of the system, and the Rabi splitting reaches the value $2g$ when the linewidths of BIC and exciton match. This shows that the general coupling rules are still valid, however these effects are negligible compared to the overall trend (**Figure S7j**). Importantly, the coupling strength itself is altered by varying the radiative Q factor, showing that our monolithic TMDC BIC platform is able to directly control the coupling conditions in the system.

From the above discussion, we deduce that the coupling strength is indeed independent of the electric near field enhancement itself and is always maximized, because of the optimal mode overlap between exciton and BIC in our monolithic metasurface architecture. Moreover, because of the great abundance of excitons and a well-defined mode volume, leading to a constant excitonic density, we discover an additional mechanism of changing the coupling strength by controlling the radiative loss channel of our monolithic

TMD BIC metasurface. While intrinsic losses change the Rabi splitting because of a linewidth broadening of the BIC, eventually turning to the weak-coupling regime, our resonator is made out of the absorbing material itself, which conserves the energy inside the cavity and keeps the coupling strength constant. On the other side, energy is dissipated from the resonator system through the radiative loss channel to the far field, which reduces the interaction time of the photons with the excitons for large asymmetry parameters, thus lowering the coupling strength.

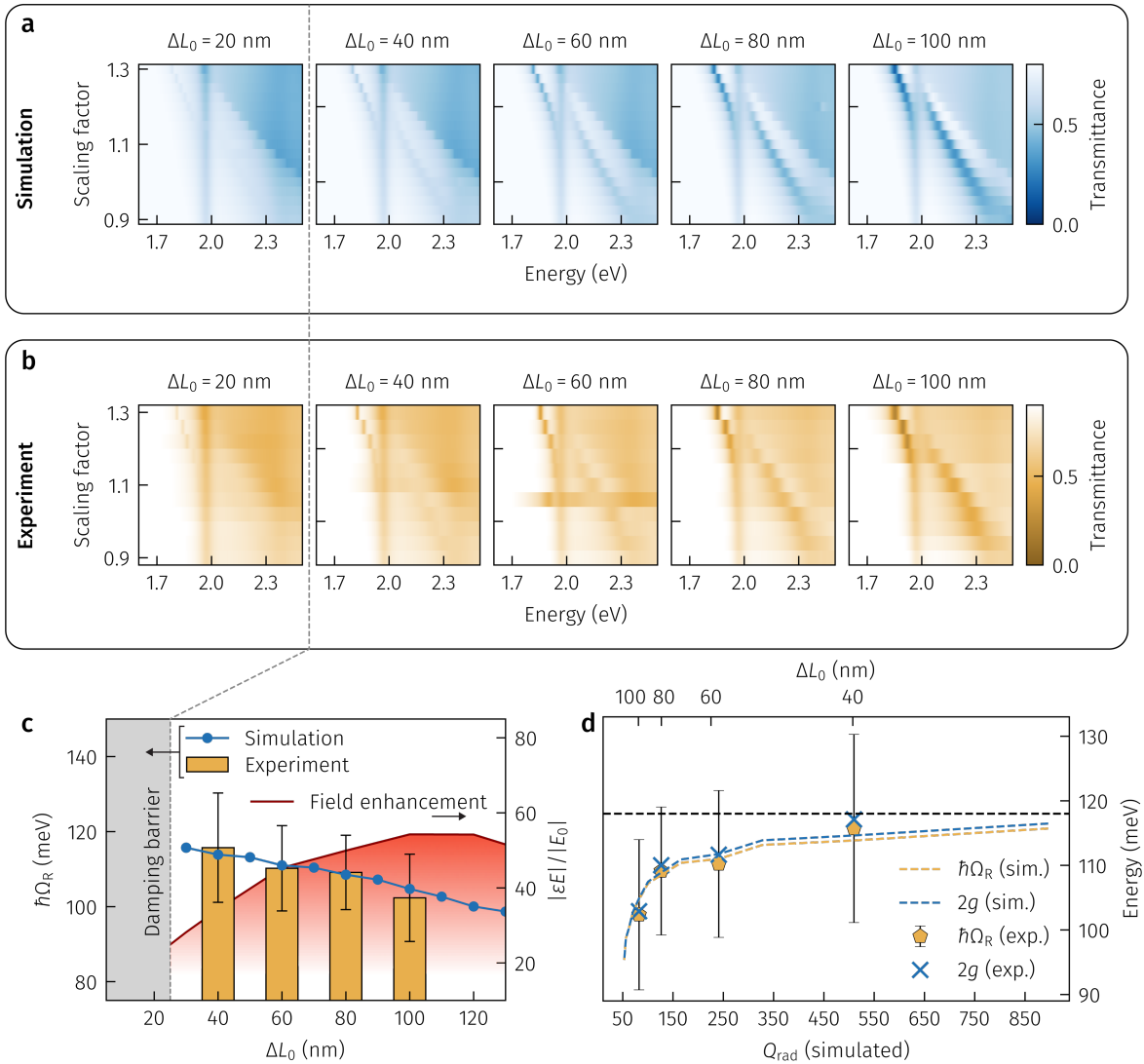

**Figure 4: Tailored Rabi splitting in BIC metasurfaces. a**, Simulated and **b**, Experimental energy dispersion plots for WS$_2$ BIC metasurfaces for different asymmetry factors $\Delta L_0$, enabling strong coupling experiments with multiple resonance quality factors and field confinements. **c**, Extracted Rabi splitting values from the data in panel (a) and averaged electric field enhancement in the resonator as a function of asymmetry parameter $\Delta L_0$. The Rabi splitting decreases with increasing asymmetry, demonstrating the capability of tailoring light-matter coupling strength using photonic BICs. **d**, Rabi splitting values and coupling constants extracted from panel (b) as a function of the simulated radiative BIC quality factor saturating for high radiative quality factors.

**Discussion**

We have experimentally realized a monolithic $WS_2$ metasurface platform based on symmetry-protected BICs. We achieved BIC-driven resonances with total Q factors of up to 370 on the low energy side of the exciton at a wavelength range from 700 to 800 nm, demonstrating that exfoliated bulk TMDCs are an excellent material for all-dielectric nanophotonics owing to their high refractive index, low losses and mono-crystallinity. Moreover, we leveraged the precise control of the resonance position of the BIC mode via geometric variations to tune the BIC spectrally to the material-intrinsic excitonic resonance at 629 nm, leading to strong light-matter interaction and hybridization of photonic and excitonic modes. We experimentally resolved a clear anti-crossing pattern of the exciton-polariton branches with a Rabi splitting of 116 meV at temperatures ranging from 5 K to room temperature. Overall, our monolithic metasurface-based approach exhibits a robust strong-coupling regime in a large range of temperatures, and even benefits from room temperature operation without compromising the signatures of strong coupling. By varying the radiative Q factor of the BIC via the asymmetry factor of the BIC unit cell between $\Delta L_0 = 40$ nm and $\Delta L_0 = 100$ nm, we were able to tune the Rabi splitting in a linear fashion, where the highest Rabi splitting value was reached for the largest Q factors. We have further shown that the Rabi splitting is independent of the electric near field enhancement due to the optimal mode overlap between BIC and exciton, rendering the coupling conditions of our monolithic TMDC BIC platform independent of the material intrinsic losses. Furthermore, because of the precise control over the radiative coupling conditions of the BIC to the far-field, we can tune the coupling strength directly, which is facilitated through a complex interplay between mode volume and temporal field confinement via the radiative loss channel. Through our generalized description of $WS_2$ as a Lorentz-type dielectric, our results are broadly applicable to other materials systems supporting resonant material-intrinsic responses such as perovskites or other van der Waals materials like black phosphorus and hexagonal boron nitride, which could pave the way for robust and tunable polaritonic devices in a large wavelength region.

**Online methods**

**Numerical Simulation**. Simulations were conducted with CST Studio Suite 2021 using periodic Floquet boundary conditions and a plane-wave excitation polarized along the long axis of the rods with normal incidence angle. The WS$_2$ material data used throughout the manuscript was adapted from tabulated permittivity data for monolayer WS$_2$ from Ref.[41] by fitting a Tauc-Lorentz model with four oscillators to represent the in-plane permittivity. We accounted for the spectral shift of the exciton and a reduction of oscillator strength for the bulk material by changing the corresponding values of the parametric material and confirmed the correctness of simulation by comparing with results from optical characterization. The exact formulas and parameters can be found in Supporting Note 1. The out-of-plane permittivity was set to $\varepsilon = 7 + 0i$.

**Nanofabrication**. WS$_2$ flakes were transferred from bulk crystals (HQ Graphene) onto fused silica substrates via mechanical exfoliation. For alignment purposes, a marker system was fabricated onto the substrates via optical lithography (SÜSS Maskaligner MA6) before exfoliation. The height of the flakes was measured with a profilometer (Bruker Dektak XT) using a stylus with a radius of 2 μm, providing sub-nanometer resolution. A layer of PMMA 950k resist followed by Espacer 300Z was spin-coated onto the sample, after which the metasurface design was written into the resist via e-beam lithography (Raith ELine Plus). A gold hardmask was deposited with e-beam evaporation followed by a lift-off in Microposit Remover 1165. Finally, the design was etched into the flakes using reactive-ion etching (Oxford PlasmaPro 100) with a SF$_6$-based chemistry at a pressure of 20 mTorr and a RF power of 50 W. The hardmask was removed in a solution of potassium monoiodide and iodine (Sigma-Aldrich).

**Optical Characterization**. The samples were characterized in a confocal optical transmission microscope with from-the-bottom illumination with collimated and linear polarized white light. The light was collected using a 50x objective with a numerical aperture of 0.8, dispersed in a spectrometer with a grating groove density of 150 mm$^{-1}$ and detected with a SI-CCD sensor.

The low-temperature transmittance measurements were performed in a variable-temperature helium flow cryostat with a confocal microscope in transmission configuration. Thermal light from a tungsten halogen light source was used for excitation and was focused from-the-bottom into the sample. The light transmitted through the sample was collected using a 20X objective with a numerical aperture of 0.4. A pinhole was used as a spatial filter in the detection path yielding a detection spot size of around 7 μm. The collected light was analyzed in a 500 mm spectrometer using a 150 grooves per mm grating and detected using a CCD sensor.

**Acknowledgements**

The authors thank Kirill Koshelev and Yuri Kivshar for helpful discussions.

This work was funded by the Deutsche Forschungsgemeinschaft (DFG, German Research Foundation) under grant numbers EXC 2089/1 – 390776260 (Germany´s Excellence Strategy), the Bavarian program Solar Energies Go Hybrid (SolTech), and the Center for NanoScience (CeNS). L. S. acknowledges funding support through a Humboldt Research Fellowship from the Alexander von Humboldt Foundation. A. B. M. gratefully acknowledges funding from the International Max Planck Research School for Quantum Science and Technology (IMPRS-QST). N. P. W. acknowledges support from the Deutsche Forschungsgemeinschaft (DFG, German Research Foundation) under Germany's Excellence Strategy—EXC-2111—390814868. S. A. M. acknowledges the funding support from the Deutsche Forschungsgemeinschaft, the EPSRC (EP/W017075/1), and the Lee-Lucas Chair in Physics. A. T. acknowledges the Emmy Noether Program TI 1063/1.

**Author contributions**

T. W., L. K and A. T. conceived the idea and planned the research. T. W., L. K. and J. K. contributed to the sample fabrication. T. W. performed optical measurements at room temperature. T. W., A. B. M. and N.P.W. performed optical measurements in the cryostat. T. W conducted the numerical simulations and data processing. T. W., A. T., L. K. and L. S. contributed to the data analysis. S. A. M. and A. T. supervised the project. All authors contributed to the writing of the manuscript.

**Competing interests**

The authors declare no competing interests.

# Strong light-matter interaction with self-hybridized bound states in the continuum in monolithic van der Waals metasurfaces

# Supplementary Information

*Thomas Weber[1], Lucca Kühner[1], Luca Sortino[1], Amine Ben Mhenni[2], Nathan P. Wilson[2], Julius Kühne[1],*

*Jonathan J. Finley[2], Stefan A. Maier[3,4,1] and Andreas Tittl[1,\**

1. Ludwig-Maximilians-Universität München, Faculty of Physics, Chair in Hybrid Nanosystems, 80539 Munich, Germany
2. Walter Schottky Institut, Department of Physics and MCQST, Technische Universität München, 85748 Garching, Germany
3. School of Physics and Astronomy, Monash University, Clayton, Victoria 3800, Australia.
4. Department of Physics, Imperial College London, London SW7 2AZ, UK

*e-mail: andreas.tittl@physik.uni-muenchen.de

**Supplementary Note 1: Parametric Tauc-Lorentz material for $WS_2$**

The imaginary part of the permittivity of a Tauc-Lorentz material is given by

$$\text{Im } \varepsilon(E) = \begin{cases} \dfrac{A_k E_{0,k} C_k (E - E_g)^2}{(E^2 - E_{0,k}^2)^2 + C_k^2 E^2} \dfrac{1}{E}, & \text{for } E > E_g \\ 0, & \text{for } E \leq E_g \end{cases} \qquad (S1)$$

where $A_k$ is the amplitude (or oscillator strength), $E_{0,k}$ the resonance position and $C_k$ the damping rate of the k-th oscillator and $E_g$ a global bandgap applied to all oscillators set to 300 THz. The real part of the permittivity was retrieved via the Kramers-Kronig relations

$$\text{Re } \varepsilon(E) = D + \frac{2}{\pi} P \int_{E_g}^{\infty} \frac{\xi \, Im(\varepsilon(\xi))}{\xi^2 - E^2} d\xi \qquad (S2)$$

where the background permittivity was set with D to 10.

| k | $A_k$ (arb. u.) | $E_{0,k}$ (THz) | $C_k$ (THz) |
|---|---|---|---|
| 1 (Exciton) | 546.9 | 476.6 | 14.0 |
| 2 | 420.5 | 578.1 | 41.1 |
| 3 | 515.4 | 683.5 | 43.1 |
| 4 | 9091.1 | 735.3 | 216.3 |

**Table S1:** Parameters for artificial $WS_2$ Tauc-Lorentz material.

**Supplementary Note 2: Complete and simplified coupled oscillator models for strong coupling**

A powerful toolkit to describe resonances in open cavities and their coupling properties to both the far field and to other modes is the temporal coupled mode theory (TCMT)[1].

Here, we utilize a model consisting of ports, which allows for transmitted and reflected waves, to accurately describe the conditions of our TMD-BIC metasurface (**Figure S1**). The system is excited via external fields through port 1 ($\mathbf{s}_+ = (s_{1+}, 0)^T$), which are subsequently affected by the metasurface, described by the scattering matrix S, and decay through port 1 and 2 with equal proportion due to mirror symmetry ($\mathbf{s}_- = (s_{1-}, s_{2-})^T$) (Equation (S1)).

$$\mathbf{s}_- = S\mathbf{s}_+ \quad (S3)$$

For the two-port system the scattering matrix S is a symmetric 2x2 matrix given by

$$S = \begin{pmatrix} s_{11} & s_{12} \\ s_{21} & s_{22} \end{pmatrix} = C + K[\mathrm{i}(\omega I - \Omega) + \Gamma]^{-1} K^T, \quad (S4)$$

where we can attribute the S-parameter $s_{11} = \frac{s_{1-}}{s_{1+}} = r$ with the reflection and $s_{21} = \frac{s_{2-}}{s_{1+}} = t$ with the transmission coefficient. The resonance frequencies and intrinsic damping rates of the modes of the cavity are given by $\Omega$ and the radiative damping rates by $\Gamma$. The port-cavity coupling is described by $K$, which is a function of the radiative damping rates. The direct port-port crosstalk by the unitary matrix $C$.

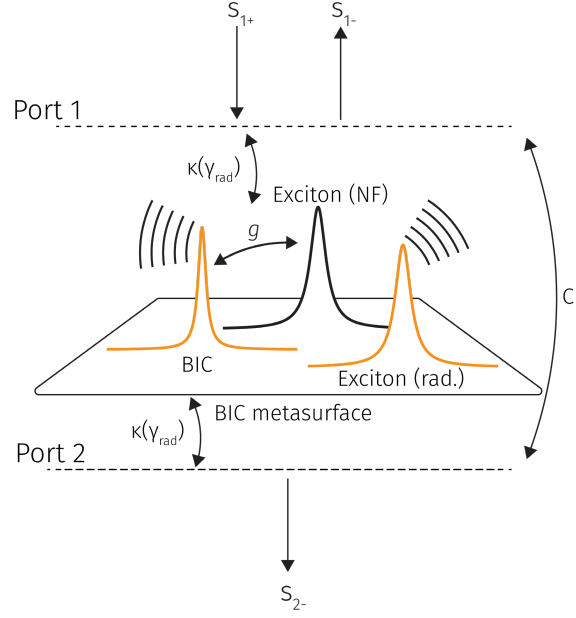

**Figure S1:** Sketch of the used TCMT model with two ports and three resonances: BIC and two excitons to describe the near-field coupling and radiative contributions. The metasurface is illuminated from above ($s_{1+}$) and allows for reflected ($s_{1-}$) and transmitted waves ($s_{2-}$). The near-field coupling between BIC and the exciton is described by the coupling parameter g, the far-field coupling via the parameter κ as a function of the radiative damping rate of the respective resonance. Direct port coupling is addressed via parameter C.

From experimental and simulated transmittance spectra with strong-coupling fingerprint, a distinct three-dip feature is visible, consisting of the coupled BIC-exciton polariton and radiative parts of the exciton itself, which do not participate in the coupling process. Thus, we can describe the system effectively by a cavity supporting three resonant modes (BIC and a two-fold contribution of the exciton), where one part only interacts with the BIC in the near-field via the coupling strength $g$ and the other part only via the radiative far-field via the term $\sqrt{\gamma_{BIC,rad}\,\gamma_{Ex,rad}}$. The complete set of parameters is given by the following:

$$\Omega = \begin{pmatrix} \omega_{BIC} + i\gamma_{BIC,int} & g & 0 \\ g & \omega_{Ex} + i\gamma_{Ex,int} & 0 \\ 0 & 0 & \omega_{Ex} + i\gamma_{Ex,int} \end{pmatrix}$$

$$\Gamma = \begin{pmatrix} \gamma_{BIC,rad} & 0 & \sqrt{\gamma_{BIC,rad}\,\gamma_{Ex,rad}} \\ 0 & 0 & 0 \\ \sqrt{\gamma_{BIC,rad}\,\gamma_{Ex,rad}} & 0 & \gamma_{Ex,rad} \end{pmatrix}$$

$$K = \begin{pmatrix} \sqrt{\gamma_{\text{BIC,rad}}} & 0 & \sqrt{\gamma_{\text{Ex,rad}}} \\ \sqrt{\gamma_{\text{BIC,rad}}} & 0 & \sqrt{\gamma_{\text{Ex,rad}}} \end{pmatrix}$$

$$C = e^{i\phi} \begin{pmatrix} r_0 & it_0 \\ it_0 & r_0 \end{pmatrix} \qquad (S5)$$

This allows to extract the coupling strength using a single spectrum. However, it is limited to simplified materials with Lorentz-type oscillators, that is, no dispersive losses due to inter-band absorption. However, we can reduce the TCMT model to a simpler eigenvalue problem, which we use to fit the dispersion of the polariton branches in the main text. By combining the radiative and intrinsic damping rates into a total damping rate $\gamma_i = \gamma_{i,\text{rad}} + \gamma_{i,\text{int}}$, we arrive at the expression

$$\begin{pmatrix} \omega_{\text{BIC}} + i\gamma_{\text{BIC}} & g \\ g & \omega_{\text{Ex}} + i\gamma_{\text{Ex}} \end{pmatrix} \mathbf{v} = \omega_{\pm} \mathbf{v}, \qquad (S6)$$

where $\omega_{\pm}$ are the resonance positions of the upper and lower polariton branches and $\mathbf{v}$ contains the Hopfield coefficients, which describe the proportions of BIC and exciton in the hybrid mode[2]. By solving the Eigenvalue problem, the well-known polariton dispersion relation

$$\omega_{\pm} = \frac{\omega_{\text{BIC}} + \omega_{\text{Ex}}}{2} + \frac{i(\gamma_{\text{BIC}} + \gamma_{\text{Ex}})}{2} \pm \sqrt{g^2 - \frac{1}{4}(\gamma_{\text{BIC}} - \gamma_{\text{Ex}} + i(\omega_{\text{BIC}} - \omega_{\text{Ex}}))^2} \qquad (S7)$$

is retrieved, which we use to fit the coupling strength $g$ from experimental results. The Rabi splitting $\Omega_R$ is defined as the minimal polariton splitting $\Delta\omega = \omega_+ - \omega_-$, i.e. when $\omega_{\text{BIC}} = \omega_{\text{Ex}}$, yielding

$$\Omega_R = 2\sqrt{g^2 - (\gamma_{\text{BIC}} - \gamma_{\text{Ex}})^2/4}. \qquad (S8)$$

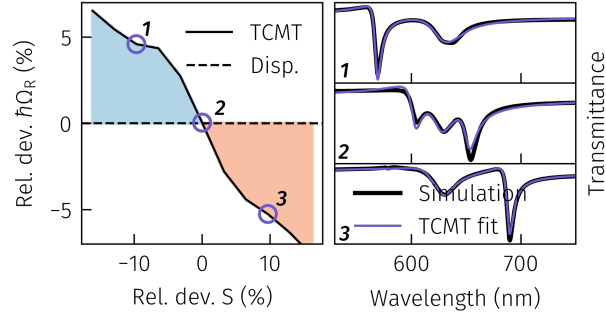

**Figure S2: Comparison between TCMT and dispersion fit for strong coupling.** The Rabi splitting extracted from the TCMT model reveals an increase of coupling strength with a reduction of the unit cell scaling factor and vice versa. The Rabi splitting extracted from the polariton dispersion model is inherently independent of the scaling factor. The comparison of the two models shows that the dispersion model yields the averaged Rabi splitting of all scaling factors and is hence a suitable model to describe strong coupling in monolithic WS$_2$ BIC metasurfaces.

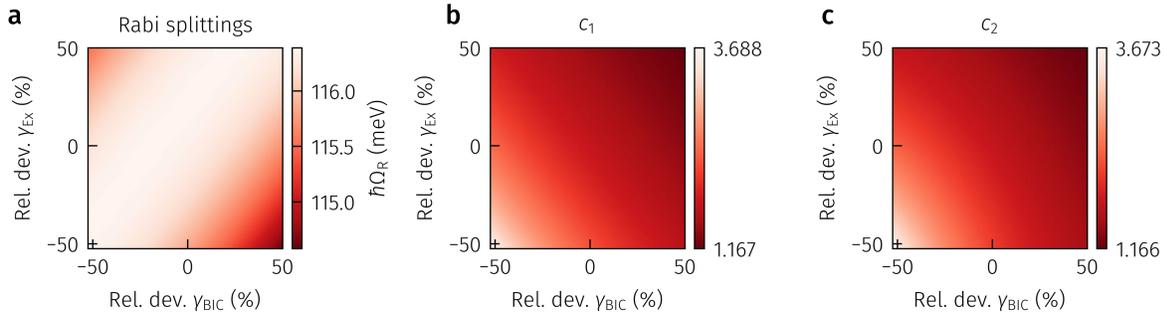

**Figure S3: Influence of uncertainties of resonance linewidths on strong coupling. a**, Fitted Rabi splittings for the data in Figure 3b of the main text for deviating linewidths of BIC and exciton. For the largest mismatch of linewidth (50% narrower exciton and 50% broader BIC), the reduction of the Rabi splitting is less than 2%, which is in the range of uncertainty of the fitted Rabi splitting for $\hbar\gamma_{BIC} = 30$ meV and $\hbar\gamma_{Ex} = 36$ meV. **b**, **c**, The strong-coupling criteria $c_1$ and $c_2$ (equations (2) and (3) of the main text) for the conditions of panel (a) show always values greater than one, proving that the monolithic TMD BIC metasurface reaches a stable point in the strong coupling regime.

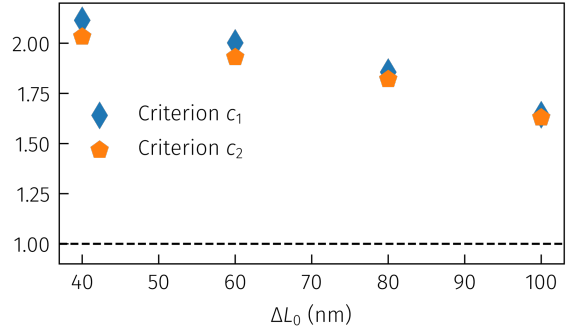

**Figure S4: Strong-coupling criteria for different BIC asymmetry parameters.** Both criteria remain well over the threshold of one, even for largest asymmetry, which shows that the system is always in the strong-coupling regime.

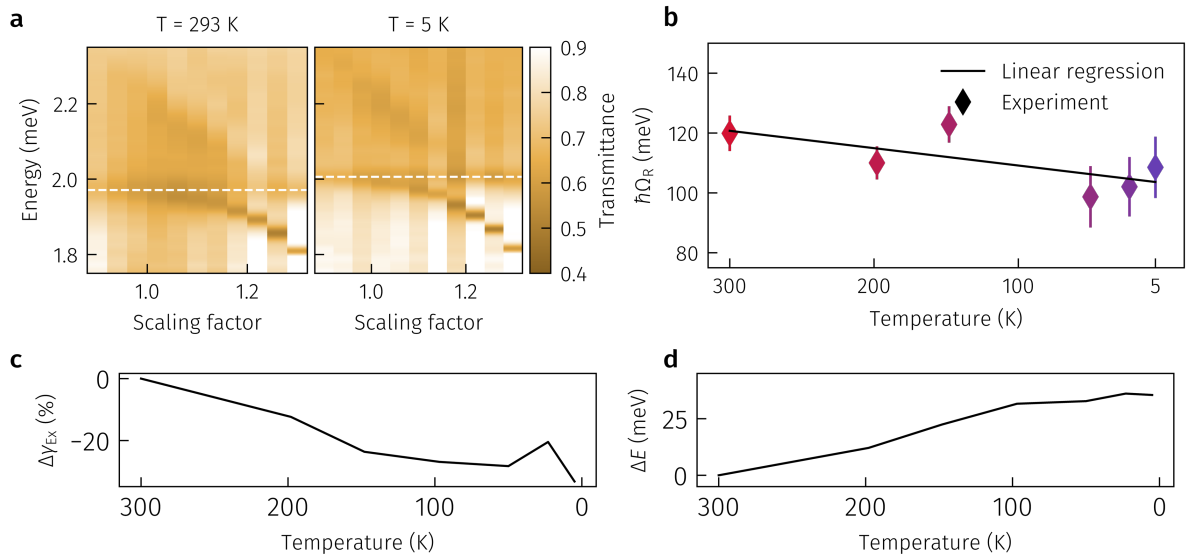

**Figure S5: Experimental results of measurements in a cryostat. a**, Energy dispersion plots of strong coupling in $WS_2$ BIC metasurfaces with an asymmetry parameter $\Delta L_0 = 80$ nm and a height of 33 nm at room temperature ($T = 293$ K) and cryogenic temperatures ($T = 5$ K) depicting a spectral blueshift of the $WS_2$ exciton. **b**, Reduced Rabi splitting for low temperatures.
**c**, Relative change of the excitonic linewidth, extracted from $WS_2$ metasurfaces with symmetric unit cell, showing a decrease for lower temperatures. **d**, Relative change of the energy of the $WS_2$ exciton, showing a spectral blueshift for lower temperatures.

**Supplementary Note 3: Effect of the radiative and intrinsic quality factors on the coupling strength and Rabi splitting**

Our monolithic WS$_2$ metasurfaces give great control over the radiative quality factor of the BIC resonances. However, the intrinsic quality factor, which is affected by material-intrinsic losses and other parasitic loss channels, such as fabrication imperfections, sets an upper bound on the total achievable quality factor, especially at low asymmetry parameters $\Delta L_0$ (Figure 2f and 3d). Furthermore, the presence of losses limits and shifts the field enhancement to the critical coupling condition. Here, we compare the effect of both the radiative BIC quality factor (**Figure S6**) by changing the symmetry parameter and the intrinsic BIC quality factor (**Figure S7**). This is achieved by artificially increasing the imaginary part of the permittivity and show that the radiative quality factor directly influences the coupling strength (and thus the Rabi splitting), whereas the intrinsic quality factor has no effect on the coupling strength and only influences the Rabi splitting by means of mismatch of the quality factors of BIC and excitonic response. Moreover, this implies that the field enhancement is not the driving force behind the strong-coupling process.

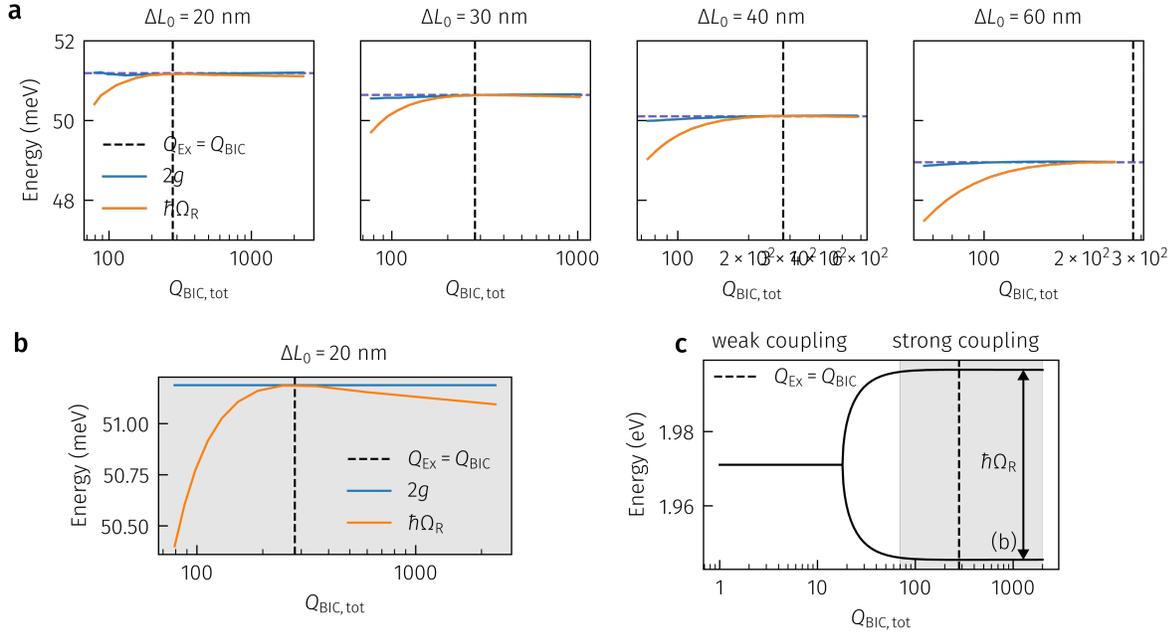

**Figure S6: Effect of the intrinsic quality factor on the coupling strength and Rabi splitting. a**, Fitted Rabi splittings and coupling strengths plotted against the total quality factor, which is varied via intrinsic loss channels (cmp. Figure S2a). The coupling strength is constant over the complete range of quality factors, whereas the Rabi splitting decreases for lower Q factors. Overall the coupling strength and Rabi splitting decreases for larger asymmetry factors, which is consistent with Figure 2. **b**, Fitted Rabi splittings and coupling strength by assuming the coupling strength is constant for $\Delta L_0 = 20$ nm. The Rabi splitting has a maximum, when the quality factor of the BIC matches the quality factor of the exciton. **c**, Mode splitting for parameters extracted from panel (b). The decrease in Rabi splitting for lower Q factors is caused by moving towards the weak coupling regime.

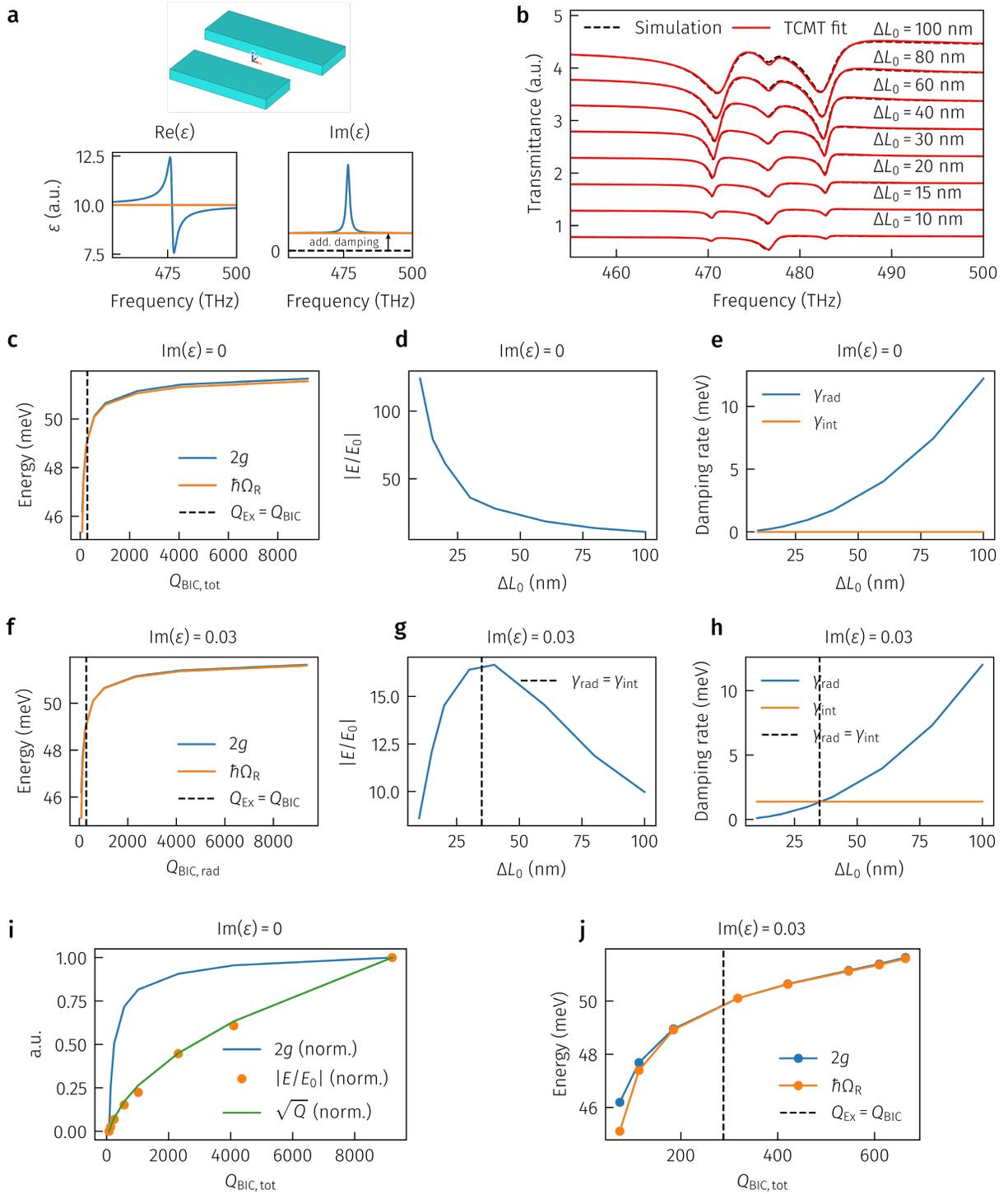

**Figure S7: Effect of radiative quality factor on coupling strength and Rabi splitting. a**, Sketch of a simplified unit cell in vacuum with a height of 33 nm. The lateral geometric parameters are taken from the main text. A simplified Lorentzian material with the constants A = 6.837318, $E_0$ = 476.6 THz, C = 1.4033 THz, $E_g$ = 0 was used, where additional absorption can be introduced via an offset in the imaginary part of the permittivity. **b**, Simulated and fitted transmittance spectra for different asymmetry parameters. The total excitonic linewidth comprising of the radiative and intrinsic parts extracted from transmittance spectra was fitted to be 0.83011 THz, yielding a Q factor $Q_{Ex}$ = 287.175612, which is without loss of generality chosen to be narrower than the $WS_2$ A-exciton, to be able to resolve features even for high asymmetry factors. **c**, The coupling strength and Rabi splitting plotted against the BIC quality factor for the aforementioned Lorentzian material without additional losses, showing only slight deviation for $Q_{BIC} \gg Q_{Ex}$. **d**, Mean field enhancement inside the resonator structure, which scales with different asymmetry parameters as $\Delta L_0^{-1}$. **e**, Intrinsic and external damping rates of the BIC without excitonic response. The radiative rate scales according to the inverse square dependence of the radiative quality factor like $\Delta L_0^2$, whereas the intrinsic damping rate is zero due to vanishing losses. **f-h**, Similar to panels (c) - (e) with intrinsic losses introduced via an offset of the imaginary part of the permittivity of 0.03. The coupling strength and Rabi splitting is within the simulation and fit uncertainties identical to the lossless case when assuming the radiative quality factor, whereas the field enhancement shows a different picture. The field enhancement is governed by critical coupling, showing a maximum for equal radiative and intrinsic damping rates. Overall the maximum field enhancement for the lossless case is around 7.5 times larger than for the lossy case. **i**, Comparison between coupling strength and mean field enhancement in resonator structure with respect to the BIC Q factor without losses. The field enhancement scales with $Q^{1/2}$ and would saturate due to the square root at larger Q factors than shown. The coupling strength is saturating earlier with a differing functional dependence. **j**, The Rabi splitting and coupling strength plotted against the total quality factor for a lossy material. The functional dependence differs due to a change in quality factors, however the Rabi splitting and coupling strength still show the same trend.

**Supplementary Note 4: Mode volume calculations**

To calculate the modal volume of a BIC resonance we used a generalized expression to account for non-negligible losses in the system[3]

$$\frac{1}{V_{eff}} = \text{Re}\left\{\frac{1}{v}\right\}, v = \frac{\langle\langle \mathbf{E}(\mathbf{r})|\mathbf{E}(\mathbf{r})\rangle\rangle}{\max(\varepsilon(\mathbf{r})|\mathbf{E}(\mathbf{r})|^2)} \quad (S9)$$

The inner product consists of a volume integral term and a flux integral to account for the varying degree of radiative loss for the BIC platform[4]

$$\langle\langle \mathbf{E}(\mathbf{r})|\mathbf{E}(\mathbf{r})\rangle\rangle = \int_V \varepsilon(\mathbf{r})|\mathbf{E}(\mathbf{r})|^2 \, d\mathbf{r} + \frac{c^2}{2\omega}\oint_{\partial V} \Phi(\mathbf{r}) \cdot d\mathbf{S} \quad (S10)$$

with

$$\Phi(\mathbf{r}) = -\frac{1}{2}\nabla E^2 - k^2 \mathbf{r} E^2 + \mathbf{r}\sum_{ij}\left(\frac{\partial E_i}{\partial x_j}\right)^2 - 2\sum_i K_i \nabla E_i$$

$$\mathbf{K}(\mathbf{r}) = (\mathbf{r}\cdot\nabla)\mathbf{E}(\mathbf{r}) \tag{S11}$$

where $k$ is the wavevector in free space and $\omega$ the complex frequency of the cavity mode.

We simulated the electric near fields in one unit cell with Lumerical, assuming a homogeneous environment and a constant permittivity of the resonator material of $\varepsilon = 10$, using a fixed mesh and periodic boundary conditions. To assure an accurate result, we increased the z-dimension of the simulation domain until the mode volume converged against a constant value for each value of the asymmetry parameter $\Delta L_0$.

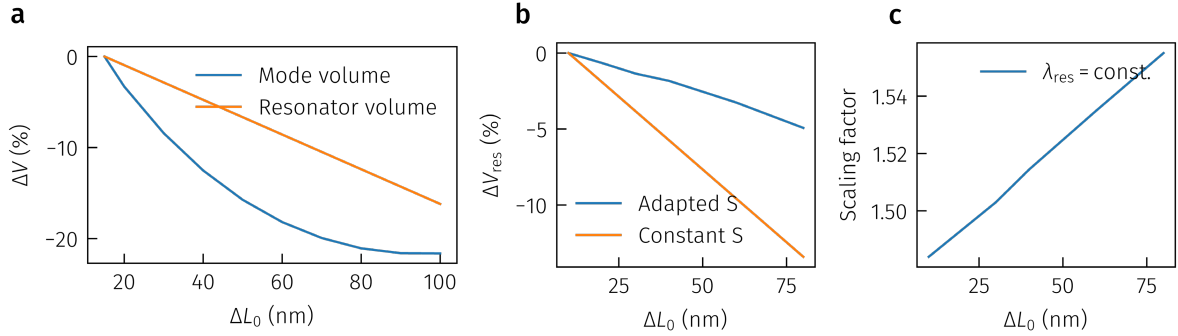

**Figure S8: Effective mode volumes in monolithic TMD BICs. a**, Relative change in effective mode volume calculated according to and resonator volume for different asymmetry factors for constant scaling factor. **b**, Relative change in the resonator volume for constant scaling factor and scaled unit cell to maintain the same BIC resonance wavelength. **c**, Unit cell scaling factor for different asymmetry parameters to maintain a constant resonance wavelength.

3. Kristensen, P. T., Vlack, C. V. & Hughes, S. Generalized effective mode volume for leaky optical cavities. *Opt Lett* 37, 1649 (2012).

4. Muljarov, E. A. & Langbein, W. Exact mode volume and Purcell factor of open optical systems. *Phys Rev B* 94, 235438 (2016).31